\def\<{\langle}
\def\>{\rangle}

\documentclass[pra,aps,twocolumn,tightenlines]{revtex4}
\usepackage{epsfig}

\begin{document}

\title{Verifying Atom Entanglement Schemes by Testing Bell's Inequality}

\author{Dimitris G. Angelakis,$^{(1)}$
Almut Beige,$^{(2)}$ Peter L. Knight,$^{(1)}$ William J.
Munro,$^{(3)}$\footnote{Present address: Hewlett-Packard
Laboratories, Filton Road, Stoke Gifford, Bristol BS34 8QZ, UK.}
and Ben Tregenna$^{(1)}$}

\address{$^{(1)}$ Optics Section, Blackett
Laboratory, Imperial College, London SW7 2BW, England. \\
$^{(2)}$ Max-Planck-Institute for Quantum Optics,
Hans-Kopfermann-Stra{\ss}e 1, 85748 Garching, Germany. \\
$^{(3)}$ Special Research Centre for Quantum Computer Technology,
The University of Queensland, 4072 Brisbane, Australia.}

\date{\today}

\begin{abstract}
Recent experiments to test Bell's inequality using entangled
photons and ions aimed at tests of basic quantum mechanical
principles. Interesting results have been obtained and many
loopholes could be closed. In this paper we want to point out that
tests of Bell's inequality also play an important role in
verifying atom entanglement  schemes. We describe as an example a scheme to
prepare arbitrary entangled states of $N$ two-level atoms using a
leaky optical cavity and a scheme to entangle atoms inside a photonic 
crystal. During the state preparation no photons are
emitted and observing a violation of Bell's inequality is the only
way to test whether a scheme works with a high precision or not.
\end{abstract}

\maketitle

{{\bf Keywords}: Bell inequality tests, atom entanglement schemes,
quantum computing, manipulation of decoherence-free states }

\section{Introduction}

Entanglement is a defining feature of quantum mechanical systems
and leads to correlations between sub-systems of a very
non-classical nature. These in turn lead to fundamentally new
interactions and applications in the field of quantum information
\cite{ben95}.
One of the main requirements for quantum computation, for instance,
is the ability to manipulate the state of a quantum mechanical system
in a controlled way \cite{Vinc}. If each logical qubit is obtained from
the states of a single atom one has to be able to
prepare {\em arbitrary entangled states} of $N$ two-level atoms.
Entangled atoms can also be used to improve frequency standards \cite{Huelga}.

The strange nature of entanglement was first pointed
out by Einstein, Podolsky and Rosen (EPR) \cite{ein35}. The deeper
mysteries of entanglement were quantified by Bell \cite{Bell65} in
his famous {\em Bell's inequalities}. These raise testable differences
between quantum mechanics and {\em all} local realistic theories.
There are a number of Bell inequalities for two subsystems where each
subsystem contains one qubit of information including the original
{\em spin} \cite{Bell65}, Clauser-Horne (CH) \cite{Clauser and Horne
74}, Clauser-Horne-Shimony-Holt (CHSH) \cite{CHSH69} and
information theoretic \cite{Braunstein and Caves 1988} Bell
inequalities, to name but a few. The concept of qubits is
thus a useful construct for Bell inequality tests, as it is
a useful construct for quantum information.

Numerous experimental tests of Bell-type inequalities have been
made using photons starting with Aspect \cite{asp82}. Violations,
showing agreement with quantum mechanics, of over 200 standard
deviations \cite{kwi95} and over large distances \cite{tit98} have
now been performed. A lot of progress has been made in closing
different loopholes. Very recently, the first experiment testing
Bell's inequality with {\em atoms} has been performed by Rowe {\em
et al.} \cite{Wineland}. The main advantage of this experiment is
that it is possible to read out the state of the atoms with a very
high efficiency following a measurement proposal by Cook
\cite{Cook,behe} based on ``electron shelving''. This allows to
investigate, characterize and test Bell's inequality with a very
high precision and to close the detection loophole
\cite{Wineland}.

In this proceedings we want to point out that tests of Bell's
inequalities do not {\em only} play an important role in testing
basic quantum mechanical principles. They are also crucial in
quantum computing to verify atom entanglement schemes. The main
obstacle in the state manipulation of atoms arises from the fact
that a quantum mechanical system is always also coupled to its
environment. This leads to decoherence. The atoms may emit
photons, and the state manipulation failed. New entanglement
schemes have therefore been based on the existence of
decoherence-free states \cite{DFS1,DFS2,DFS3,Guo98,ent} and ideas
of how to restrict the time evolution of a system onto the
corresponding decoherence-free subspace \cite{ent,Tregenna,kempe}.
But during the state preparation no photons are emitted and
observing a violation of Bell's inequality is the only way to test
whether the scheme works with a high precision or not.

This paper is organized as follows. In the next section we 
discuss a scheme based on recent results by Beige {\em et al.} 
\cite{ent,Tregenna} which allow to prepare $N$ two-level atoms, 
each with a state $|0\>$ and $|1\>$, in an arbitrary entangled 
state by using a leaky optical cavity. Section III discusses a recent 
proposal to entangle atoms in a photonic band gap.
Section IV illustrates different methods to verify highly entangled 
states for a few atoms. We conclude in Section V.

\section{The Preparation of arbitrary entangled states}

\begin{figure}[!ht]
\epsfig{file=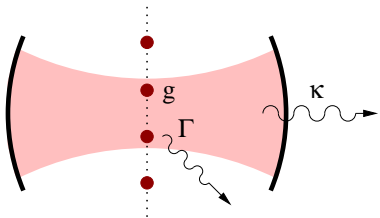,width=5cm} \caption{Experimental setup
for the preparation of arbitrary entangled states. The system
consists of two two-level atoms which can be moved in and out of
an optical leaky cavity. An atom inside the cavity couples to the
cavity mode with coupling strength $g$. A photon leaks out through
the cavity mirrors with a rate $\kappa$ and $\Gamma$ is the
spontaneous decay rate of each atom.}\label{cavity}
\end{figure}

In this Section we want to give a short overview about a scheme to
prepare $N$ two-level atoms in an {\em arbitrary} entangled state
by using a {\em leaky optical cavity} as shown in Fig.
\ref{cavity}. The atoms can be stored in a linear trap, an optical
lattice or on a chip for quantum computing
\cite{haensch,schmiedmeyer} and it is possible to move two of them
simultaneously into the cavity. The coupling constant of each atom
inside the cavity to the single field mode is in the following
denoted by $g$. The spontaneous decay rate of each atoms equals
$\Gamma$, while $\kappa$ is the decay rate of a single photon
inside the cavity.

\subsection{The preparation of an entangled state of two atoms}

To explain the main idea of the entanglement scheme we consider first
only {\em two} two-level atoms with the states $|0\>$ and
$|1\>$ which are in resonance with the cavity mode, 
$\omega_{\rm cav}=\omega_0$, and describe the preparation of the
entangled state 
\begin{eqnarray} \label{instate}
|\varphi \> &=& \alpha \, |a\> + \sqrt{1-|\alpha|^2} \, |00\> ~,
\end{eqnarray}
where
\begin{eqnarray} \label{a}
|a\> & \equiv & \big( |10\> - |01\> \big)/\sqrt{2} 
\end{eqnarray}
is a maximally entangled state of the two atoms 
and $\alpha$ an arbitrary parameter.
To do so the two atoms, initially in the ground state, $|\psi_0\> = |00\>$,
have to be moved into the cavity which should be empty. In
addition, a single laser pulse of length $T$, which is in
resonance with the atoms, $\omega_{\rm laser}=\omega_0$, has to
be applied. This laser couples with a Rabi frequency
$\Omega^{(i)}$ to atom and we define
\begin{eqnarray} \label{omega-}
\Omega^{(-)} &\equiv& \big(\Omega^{(1)}-\Omega^{(2)}\big)/\sqrt{2} ~.
\end{eqnarray}

There are two sources of decoherence in the system: spontaneous
emission by each atom with rate $\Gamma$ and leakage of photons
through the cavity mirrors with rate $\kappa$. We assume here that
the later one is the main source for dissipation and a strong
coupling between the atoms and the cavity mode, i.e.
\begin{equation} \label{par}
\Gamma \ll |\Omega^{(-)}| \ll g^2/\kappa ~{\rm and}~ \kappa ~.
\end{equation}
For this parameter regime, one can show that the state of the
atoms at the end of the pulse is given by Eq. (\ref{instate}) with
\begin{eqnarray} \label{T}
\alpha &=& - {\rm i} \, {\Omega^{(-)} \over |\Omega^{(-)}|} \,
\sin \left( {|\Omega^{(-)}| T \over 2} \right) ~.
\end{eqnarray}
To do so we now have a closer look at the time evolution of the
system.

To describe the time evolution of the system we make use of the
quantum jump approach \cite{HeWi1,HeWi2,HeWi3}. It predicts
that the (unnormalized) state of a quantum mechanical system under
the condition of no photon emission equals at time $t$
\begin{eqnarray} \label{phit}
|\psi^0 (t) \rangle &=& \exp \left( -{{\rm i} \over \hbar} H_{\rm
cond} t \right) |\psi_0 \rangle
\end{eqnarray}
where the conditional Hamiltonian $H_{\rm cond}$ is a
non-Hermitian Hamiltonian. For the total system consisting of the
two atoms and the cavity mode it is given by
\begin{eqnarray} \label{HcondL}
H_{\rm cond} &=&  {\rm i} \hbar g \, \sum_{i=1}^2 \big(
b \, |2\>_{ii} \<0| - {\rm h.c.} \big) \nonumber \\
& & + {\hbar \over 2} \sum_{i=1}^2 \big( \, \Omega^{(i)} \,
|2\>_{ii}\<0| + {\rm h.c.} \, \big) \nonumber \\
& &  - {\rm i} \hbar \Gamma \sum_{i=1}^2 |2\>_{ii} \<2| - {\rm i}
\hbar \kappa \, b^\dagger b ~,
\end{eqnarray}
where $b$ is the annihilation operator for a single photon in the cavity mode.
As a consequence of the non-Hermitian terms in Eq. (\ref{HcondL}), the
norm of the state vector $|\psi^0 (t) \rangle$ decreases with time
and its squared norm at time $t$ gives the probability for no
photon emission in $(0,t)$, i.e.
\begin{eqnarray} \label{P0}
P_0(t) &=& \| \, |\psi^0 (t) \rangle \, \|^2.
\end{eqnarray}
For the parameter choice (\ref{par}) we can show that this
probability is very close to unity, which is why we have to
consider only the no-photon time evolution of the system.

\begin{figure}[!ht]
\epsfig{file=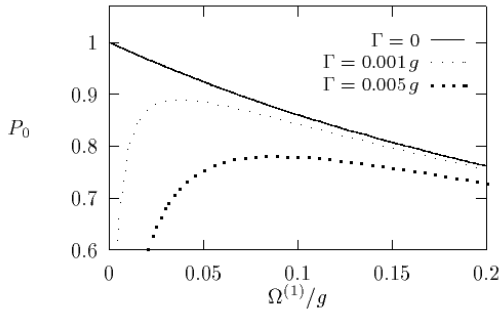,width=7cm}
 \caption{The probability for no
photon emission during the preparation of the maximally entangled
state $|a\>$ for different Rabi frequencies $\Omega^{(1)}$ and
$\Omega^{(1)}=-\Omega^{(2)}$, different spontaneous decay rates
$\Gamma$ and $\kappa=g$.}\label{fig6}
\end{figure}

Because of Eq. (\ref{par}), there are two very different time
scales in the system. One can therefore solve the conditional time
evolution of the system with the help of an adiabatic elimination
of the fast varying states. This has been done in Ref. \cite{ent}.
Alternatively one can solve Eq. (\ref{phit}) numerically. Here we
only present numerical results. Fig.~\ref{fig6} shows the
probability for no photon emission obtained from a numerical
integration of Eq.~(\ref{phit}). As an example we choose the
preparation of the maximally entangled state $|a\>$ with
$T=\pi/|\Omega^{(-)}|$. If a photon is emitted in $(0,T)$ the
preparation failed and has to be repeated.
Nevertheless, as Fig. \ref{fig6} shows, the success rate of the
scheme is very close to unity. 
The fidelity of the state $|\psi^0(T)\>$ in case of
a successful preparation is found to be always higher than
$95\,\%$ for the parameters chosen in Fig.~\ref{fig6}.

The main limiting factor of the scheme is
spontaneous emission by the atoms with rate $\Gamma$. To suppress
this one can make use of an additional atomic level 2. As
described in detail in Ref. \cite{Bell}, this level should form
together with the levels 0 and 1 a $\Lambda$ configuration.
Choosing the parameters as in Ref. \cite{Bell}, level 2 can be
eliminated adiabatically, which leads to atomic two-level systems
as described above. But now the states $|0\>$ and $|1\>$ are
stable ground states and do not decay.

We now want to give a more intuitive explanation why the scheme
works. It takes advantage of the fact that two-level atoms inside
the cavity possess {\em decoherence-free} states
\cite{ent,DFS1,DFS2,DFS3}. Decoherence-free states arise if no
interaction between the system and its environment of free
radiation fields takes place. If we neglect spontaneous emissions
this is exactly the case if the cavity mode is empty
\cite{ent}. In addition, the systems own time evolution due to the
interaction between the atoms and the cavity mode should not lead
to the population of non decoherence-free states. From this
condition we find that the decoherence-free states are the
superpositions of the two atomic states $|00\>$ and $|a\>$ while
the cavity mode is empty. Eq. (\ref{instate})
corresponds therefore to the decoherence-free states of the
system.

How the preparation scheme works can now easily be understood in
terms of an {\em environment induced quantum Zeno effect}
\cite{misra,Itano,zeno}. Let us define $\Delta T$ as the time in
which a photon leaks out through the cavity mirrors with a
probability very close to unity if the system is initially
prepared in a state with no overlap with a decoherence-free state.
On the other hand, a system in a decoherence-free state will
definitely not emit a photon in $\Delta T$. Therefore the
observation of the free radiation field over a time interval
$\Delta T$ can be interpreted as a measurement of whether the
system is decoherence-free or not \cite{messung}. The outcome of
the measurement is indicated by an emission or no emission of a
photon. As it has been shown in Ref.~\cite{messung}, $\Delta T$ is
of the order $1/\kappa$ and $\kappa/g^2$ and much smaller than the
typical time scale for the laser interaction.

Here the system continuously interacts with its environment and
the system behaves like a system under continuous observation.
Therefore, the quantum Zeno effect predicts that
all transitions to non decoherence-free states are strongly
inhibited. Nevertheless, there is no inhibition of transitions
between decoherence-free states and the effect of the laser field on
the atomic states can be described by the {\em
effective} Hamiltonian \cite{ent}
\begin{eqnarray} \label{Heff}
H_{\rm eff} &=& I\!\!P_{\rm DFS} \, H_{\rm cond} \, I\!\!P_{\rm
DFS} ~,
\end{eqnarray}
where $I\!\!P_{\rm DFS}$ is the projector on the
decoherence-free subspace. If we neglect spontaneous emission
Eq. (\ref{HcondL}) leads to
\begin{eqnarray} \label{Heff2}
H_{\rm eff}^{\rm (atoms)} &=& {\hbar \Omega^{(-)} \over 2} \,
|a\>\<00| + {\rm h.c.}
\end{eqnarray}
for the atoms, while the cavity remains empty. By solving the
corresponding time evolution, one finds that a laser pulse of
length $T$ prepares the atoms indeed in state (\ref{instate}) with
$\alpha$ given by Eq. (\ref{T}). Varying the length of the laser
pulse allows to change arbitrarily the amount of entanglement of
the prepared state.

\subsection{Generalization to $N$ atoms and arbitrary entangled states}

The entanglement scheme described in the previous subsection
allows only for the preparation of certain entangled states 
of two atoms \cite{Bell}. For many
applications, however, one has to be able to prepare arbitrary
entangled states of $N$ two-level atoms with $N \ge 2$. In this subsection we
show how this can be done by generalizing the scheme described
above.

\begin{figure}[!ht]
\epsfig{file=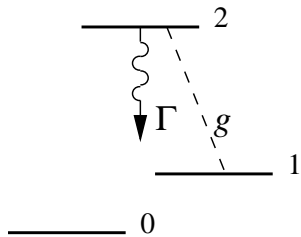,width=4cm} 
\caption{Each qubit is obtained from the ground states 
$|0\>$ and $|1\>$ of one atom. To entangle the atoms an 
additional level 2 is used which is only populated during a gate operation. 
The 1-2 transition is in resonance with the
cavity mode and couples to the cavity field with a coupling
strength $g$. The spontaneous decay rate of level 2 is denoted by
$\Gamma$.}\label{lambda}
\end{figure}

One possibility to increase the number of decoherence-free states in the
system is to move $N$ two-level atoms simultaneously into the
cavity. This has been discussed in Ref. \cite{ent} where 
the decoherence-free states of this system have been
constructed explicitly. However, for applications like quantum
computing it is crucial to have simple qubits. {\em Ideally, the
logical qubits should be the same as the physical qubits.} This can
be achieved by using three-level atoms with a $\Lambda$
configuration as shown in Fig. \ref{lambda}. In the following,
each qubit is obtained from the ground states $|0\>$ and $|1\>$ of
one atom which are the ground states in the $\Lambda$
configuration.

The setup we use is the same as shown above in Fig. \ref{cavity}.
We assume again that the atoms are trapped in a way that two of
them can be moved simultaneously into the cavity. As shown in Fig.
\ref{lambda}, the 1-2 transition is in resonance with the cavity
field, while the 0-2 transition is highly detuned and we denote
the coupling constant of each atom to the cavity mode again by
$g$. The spontaneous decay rate of level 2 is given by $\Gamma$.
Note, that the system is, in principle, scalable. Arbitrary many
atoms can be used.

To prove that it is possible to prepare arbitrary states of the
atoms we only have to show that it is possible to perform universal 
quantum gates between all qubits \cite{Vinc}. The pair of operations
we choose here and discuss in some detail in the next two
subsections are the same as discussed in Ref. \cite{Tregenna}: the
single qubit rotation and the CNOT operation. Both gates can be
realized with the help of a single laser pulse.

{\em The single qubit rotation}. The single qubit rotation is
defined by the single atom operator
\begin{eqnarray}
U_{\rm SQR}(\xi, \phi) \equiv \cos\xi-{\rm i} \sin \xi \left({\rm
e}^{{\rm i}\phi} |0\rangle \langle 1| +{\rm h.c.} \right) ~.
\end{eqnarray}
To perform this operation one can make use of an adiabatic
population transfer \cite{Vit} --- a technique which has been
realized in many experiments \cite{pop}. In order to do this, the
atom has to be moved out of the cavity. Then two lasers, both with
the same detuning, are applied simultaneously. One couples to the
1-2 transition and the other two the 0-1 transition, both with the same
Rabi frequencies. A detailed description of this process can be
found in Refs. \cite{Bell,Tregenna}.

{\em The CNOT gate}. A CNOT operation changes the state of the
target atom, which we call in the following atom 2, conditional on
the state of the control atom, atom 1, being in $|1\>$. It can be
described by the operator
\begin{eqnarray} \label{CNOT}
U_{\rm CNOT} &\equiv& |10\rangle \langle 11| + |11 \rangle \langle
10|~.
\end{eqnarray}
To perform this gate operation both atoms have to be moved into
the cavity and two laser fields are applied simultaneously. One of
the lasers couples with the Rabi frequency $\Omega_1^{(1)}$ to the
1-2 transition in atom 1, while the other excites the 0-2
transition of atom 2 with the Rabi frequency $\Omega_0^{(2)}$. In
the following we assume
\begin{eqnarray}
\Omega_1^{(1)} = \Omega_2^{(0)} \equiv \sqrt{2} \, \Omega
\end{eqnarray}
and
\begin{eqnarray}
\Gamma \ll |\Omega| \ll g^2/\kappa ~&{\rm and}&~ \kappa ~.
\end{eqnarray}
Then, as above, spontaneous emission by the atoms during the gate
operation can be neglected. In addition, it can be shown that in
this parameter regime non decoherence-free states lead to photon
emissions on a time scale much smaller than the typical time scale
for the laser interaction. As shown in Refs. \cite{Tregenna,Tregenna2}, 
the effect of the laser fields over a time
\begin{eqnarray} \label{T2}
T &=& \sqrt{2}\pi/|\Omega|
\end{eqnarray}
assembles the CNOT operation to a very good approximation.

To understand why this is the case let us first discuss what the
decoherence-free states of the system are. If we consider only the
two atoms inside the cavity and apply the same criteria as above we find
that the decoherence-free states of the system are the
superpositions of the states $|00\rangle$, $|01\rangle$,
$|10\rangle$, $|11\rangle$, which form the qubits, and
\begin{eqnarray} \label{a2}
|a \rangle &\equiv& \big( |12\rangle -|21\rangle \big)/\sqrt{2}
\end{eqnarray}
if one sets $\Omega \equiv 0$ and $\Gamma \equiv 0$. At the same
time the cavity has to be empty. 

The additional decoherence-free state (\ref{a2}) is a maximally entangled
state of the two atoms inside the cavity. Populating this state 
can therefore be used to create entanglement. 
To make sure that the laser pulse does not populate other states 
than the decoherence-free ones the scheme utilizes as above  
the environment induced quantum Zeno effect \cite{Tregenna}. 

The conditional Hamiltonian of the two atoms inside the cavity
equals here \cite{Tregenna2}
\begin{eqnarray}\label{hcond}
H_{\rm cond} &=& {\rm i} \hbar g \sum_i  \big( b \, |2\>_{ii} \<1|
- {\rm h.c.} \big) \nonumber \\
& & + {1\over 2} \, \sqrt{2} \, \hbar \Omega \,
\big( |0\>_{22} \<2| + |1\>_{11} \<2| + {\rm h.c.} \big) \nonumber \\
& & - {\rm i} \hbar \kappa \, b^{\dagger} b - {\rm i} \hbar \Gamma
\sum_i |2\>_{ii} \<2| ~.
\end{eqnarray}
Using Eq. (\ref{Heff}) this leads to the effective Hamiltonian
\begin{eqnarray}
H^{(\rm atoms)}_{\rm eff} &=& {\hbar \Omega \over 2} \, \big(
|10\rangle \langle a| - |a \rangle \langle 11| + {\rm h.c.}  \big)
\end{eqnarray}
for the atoms, while the cavity remains again empty during the gate
operation. By solving the corresponding time evolution of this
Hamiltonian one can show that a laser pulse of length $T$ as 
in Eq. (\ref{T2}) leads indeed to the realization of a CNOT gate.

\begin{figure}[!ht]
\epsfig{file=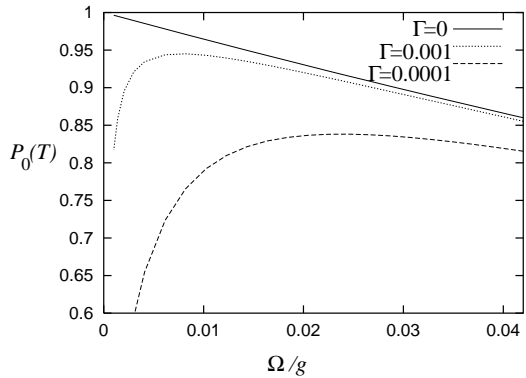,width=7cm} \caption{The probability for
no photon emission during a CNOT operation as a function of the
Rabi frequency $\Omega$, $\kappa=g$, different spontaneous
decay rates $\Gamma$ and the initial state 
$|10\>$.} \label{xxx}
\end{figure}

Finally we conclude with some numerical results. Fig.
\ref{xxx} results from a numerical integration of the time
evolution given by Eq. (\ref{hcond}) and shows the success rate of a 
single gate operation for different parameters and the initial state $|10\>$.

\begin{figure}[!ht]
\epsfig{file=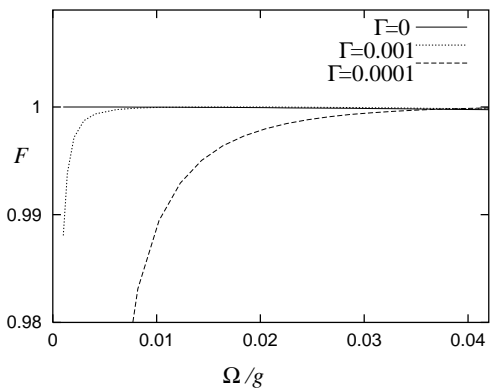,width=7cm} \caption{The fidelity for 
a single CNOT operation as a function of the Rabi frequency $\Omega$,
$\kappa=g$, different spontaneous decay rates $\Gamma$
and initial state $|10\>$.} \label{xxxx}
\end{figure}

Fig. \ref{xxxx} shows the fidelity of the prepared state. As in
the previous subsection, the fidelity of the scheme in case of no
photon emission is very high. Problems arise again from the
presence of spontaneous emission, but again this can be suppressed
as well by making use of additional levels \cite{Tregenna2}.

\section{Entangling atoms inside a Photonic crystal}

\begin{figure}[!ht]
\epsfig{figure=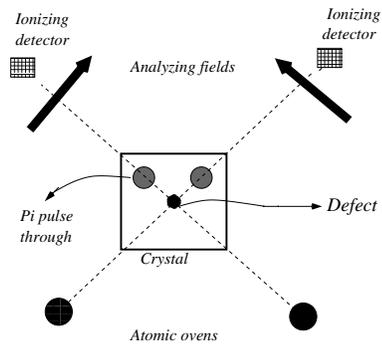,width=5.0cm} \caption{Experimental setup of the 
proposed PBG atom entanglement scheme.} \label{figx}
\end{figure}

In this section we give a short overview over another recent 
scheme by Angelakis {\em et al.} \cite{Angelakis01} to prepare two 
atoms in an entangled state using a photonic crystal
(or photonic band gap material-PG). 
The experimental setup of this scheme is shown in Fig. \ref{figx}
and the entanglement between the atoms originates from the interaction of
two atoms with a resonant defect mode.

The first atom is prepared in the upper of two optically separated
states, denoted in the following by $|e_{1}\rangle$
and propagates into the transition in
the PBG. This atom is unable to emit a photon \cite{Angelakis01}
due to the exclusion of electromagnetic modes over a continuous
range of frequencies \cite{Yablonovitch87} and hence a long lived
photon-atom bound state of the form \cite{John91,Lin99}
\begin{eqnarray}
| \Psi \rangle &=& c_{e}(t_{1}) \, |e_1\rangle|0\rangle +
c_{g}(t_{1}) \, |g_1\rangle|1\rangle,
\end{eqnarray}
is formed, where $t_{1}$ is the interaction time of atom 1 with the
defect and $|g_1\>$ the ground state of the atom. 
The coefficients $c_{e}(t_{1})$ and $c_{g}(t_{1})$ 
can be derived using a Jaynes-Cummings Hamiltonian. As soon as
atom 1 leaves the defect region, a second atom, atom 2, prepared in its
ground state $|g_2\>$ is
send through. If $t_{2}$ is the interaction time of this atom with the 
defect, then the final state of the system is of the form
\begin{eqnarray}
| \Psi_{f} \rangle
&=&c_{e}(t_{1}) \, |e_{1}\rangle|g_{2}\rangle|0\rangle+
c_{g}(t_{1})c_{e}(t_{2}) \, |g_{1}\rangle|g_{2}\rangle|1\rangle  \nonumber \\
& & + c_{g}(t_{1})c_{g}(t_{2}) \, |g_{1}\rangle|e_{2}\rangle|0\rangle
\end{eqnarray}
With an appropriate choice of $t_1$ and $t_2$ (for details see Ref.
\cite{Angelakis01}) one can achieve that the prepared states equals
\begin{eqnarray}
| \Psi_{f} \rangle&=& {1\over \sqrt 2} \, \big(
|e_{1}\rangle|g_{2}\rangle+ |g_{1}\rangle |e_{2}\rangle \big)
\otimes |0\rangle 
\end{eqnarray}
which is a maximally entangled state of the two atoms.

\section{Entanglement Verification}

One of the main problems with entanglement schemes using the
no-photon time evolution of a system is, that someone who performs
the experiment cannot know for sure whether the scheme worked or
not. Measuring whether no photon has been emitted during the
whole experiment would require perfect photon detectors which
cover all solid angels. It is therefore crucial to have other
tests whether a preparation schemes succeeds or not. 

What one can do is to use the scheme to prepare a certain maximally 
entangled state of the atoms and to verify it by observing a violation 
of a Bell inequality.
In this section we describe how to observe a violation of a Bell's 
inequality for {\em two} atoms and Mermin's GHZ inequality \cite{mermin} 
for {\em multiple} atoms. It is important to note that for each pure 
two qubit state, there exists always a Bell inequality which can 
be violated.

To do so one has to be able to
measure whether an atom is in state $|0 \>$ or $|1 \>$ with a very
high precision. This
can be done following a proposal by Cook \cite{Cook}. A detailed
analysis of this measurement scheme can be found in Ref.
\cite{behe}.

\subsection{A test for two atoms}

Given that the state (\ref{instate}) can be generated, the next
interesting question is whether it violates one of
Bell's inequalities? For certain parameters it must but what
physical measurements are necessary?

The {\it spin} (or correlation function) Bell
inequality \cite{Bell65,CHSH69} may be written formally as
\begin{eqnarray}\label{spinx}
B_{\rm S} &=& |E\left(\theta_{1},\theta_{2}\right)
-E\left(\theta_{1},\theta_{2}'\right) \nonumber \\
& & + E\left(\theta_{1}',\theta_{2}\right)
+ E\left(\theta_{1}',\theta_{2}'\right)|\leq 2~,
\end{eqnarray}
where the correlation function $E\left(\theta_{1},\theta_{2}\right)$
is given by
\begin{eqnarray}\label{correlation}
E\left(\theta_{1},\theta_{2}\right)&=&\< \sigma^{(1)}_{\theta_{1}}
\sigma^{(2)}_{\theta_{2}}\> ~.
\end{eqnarray}
Here $\theta_1$ and $\theta_2$ are real parameters.
In the following the operator $\sigma_a^{(i)}$ with $a=x$, $y$ or $z$
is the $a$ Pauli spin operators for the two-level system of atom $i$
and the operator $\sigma^{(i)}_{\theta_{i}}$ is defined as
\begin{equation} \label{yyy}
\sigma^{(i)}_{\theta_{i}} \equiv \cos \theta_{i} \,
\sigma^{(i)}_x + \sin \theta_{i} \, \sigma^{(i)}_y~.
\end{equation}

For the state considered here the correlation function depends
only on the difference between the angles $\theta_{1}$ and
$\theta_{2}$ and we have $E\left(\theta_{1},\theta_{2}\right) =
E\left(\theta_{1}-\theta_{2},0 \right)$. Now if we choose
$\vartheta = \theta_{1}-\theta_{2}=\theta_{2}-\theta_{1}'=
\theta_{1}'-\theta_{2}'$ and $\theta_{1}-\theta_{2}'=3\vartheta$
the inequality (\ref{spinx}) simplifies to
\begin{eqnarray} \label{bs}
B_{\rm S}=|3 E\left(\vartheta,0\right) - E\left(3
\vartheta,0\right)|\leq 2
\end{eqnarray}
and a violation of this inequality corresponds to $|B_{\rm{S}}|>2$.

\begin{figure}[!ht]
\epsfig{file=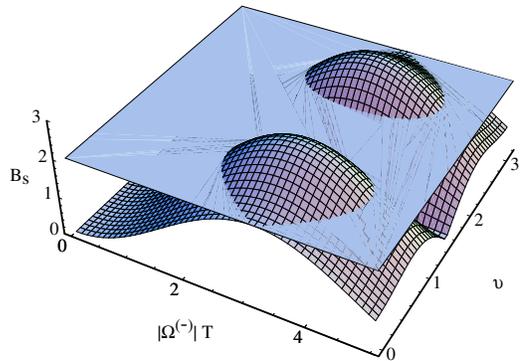,width=7cm}
 \caption{Plot of
$|B_{\rm S}|$ versus $|\Omega^{(-)}|T$ and $\vartheta$. A
violation of the {\it spin} Bell's inequality occurs for $|B_{\rm
S}|>2$ and are displayed as {\it Islands} in the $|\Omega^{(-)}|T$
- $\vartheta$ plane. The angles have been chosen so as to maximize
the violation utilizing the maximally entangled state.}
\label{fig4}
\end{figure}

It is straightforward to show that the correlation function for
the entangled state (\ref{instate}) is given by
\begin{eqnarray}
E\left(\vartheta,0\right)= - |\alpha|^{2} \cos \vartheta
\end{eqnarray}
and hence Eq.~(\ref{bs}) can assume a maximum of $|B_{\rm S}| = 2
\sqrt{2} \, |\alpha|^{2}$ for $\vartheta=\pi/4$.
Therefore, a violation of the {\it spin} Bell inequality is
possible for $|\alpha|^{2} > 1/ \sqrt 2$ given our analyzer
choices. How $|\alpha|^2$ can be expressed in terms of
the system parameter $|\Omega^{(-)}| T$ is shown in Eq. (\ref{T}). 
In Fig. (\ref{fig4}) we plot $|B_{\rm S}|$ versus $|\Omega^{(-)}| T$ and
$\vartheta$. 

A significant region of violation is observed with the maximum of
$|B_{\rm S}|=2\sqrt{2}$ occurring at $|\Omega^{(-)}|T = \pi$. The
state of the atoms at such a time is a maximally entangled state $|a\>$.
This test on Bell's inequality can therefore be used to verify the
preparation of this state with a high precision and
should be feasible with current technology. 

\subsection{A Test for multiple atoms}

So far, we have discussed how our preparation method can be tested
for two atoms. A similar procedure can be used for multiple atoms
depending on the state prepared.  For instance in the case of
three atoms we could prepare a state of the form
\begin{eqnarray}\label{spintriplet}
| \Psi \rangle ={1\over \sqrt{2}} \, \big(
| 0 \rangle |0 \rangle | 0 \rangle
+  | 1 \rangle |1 \rangle | 1\rangle \big)~.
\end{eqnarray}
This state is known as the GHZ state \cite{GHZ}. 
To characterize it we now use Mermin GHZ inequality \cite{mermin}
instead of a Bell inequality.
This inequality has the form
\begin{eqnarray}\label{oldinequality}
F &=& \left| \<\sigma_{x_{1}} \sigma_{x_{2}}\right.
\sigma_{x_{3}}\> - \<\sigma_{y_{1}} S\sigma_{y_{2}}
\sigma_{x_{3}}\> \nonumber \\
& & -\left. \<\sigma_{y_{1}} \sigma_{x_{2}}
\sigma_{y_{3}}\>-\<\sigma_{x_{1}} \sigma_{y_{2}}
\sigma_{y_{3}}\>\right| \leq 2.
\end{eqnarray}
For the state (\ref{spintriplet}) we can calculate the moments
$\<\sigma_{x_{1}} \sigma_{x_{2}} \sigma_{x_{3}}\>$,
$\<\sigma_{y_{1}} S\sigma_{y_{2}} \sigma_{x_{3}}\>$,
$\<\sigma_{y_{1}} \sigma_{x_{2}} \sigma_{y_{3}}\>$ and
$\<\sigma_{x_{1}} \sigma_{y_{2}} \sigma_{y_{3}}\>$ using a similar
procedure to that discussed in the Bell test. We find that $F=4$
for the state (\ref{spintriplet}) while the maximum according to
local realism is $F=2$. For entangled states of more atoms ($N>3$) 
a generalized form of Mermin's inequality can be employed.

\section{Discussion}

In this proceeding we used the recent results of Refs. \cite{ent,Tregenna}
and described a scheme to prepare arbitrary entangled states of $N$ 
atoms in a controlled way with a very high success rate. 
The scheme is based on the existence of decoherence-free states and 
an environment induced quantum Zeno effect to avoid the population of 
non decoherence-free states during the preparation.
In addition, we gave a short overview over a recently proposed scheme 
by Angelakis {\em et al.} \cite{Angelakis01}
to entangle two atoms inside a photonic crystal.

During the state preparation no photons
are emitted and observing a violation of Bell's inequality is one
of the ways to test whether the scheme has worked with a high
precision or not. We describe the possible violation of Bell's inequality
for two atoms. An entangled state of $N$ two-level atoms can characterized
and verified similarly with Mermin's inequality. 

{\em Acknowledgments}.
We acknowledge the support of the UK Engineering and Physical 
Sciences Research Council and the European Union and D. G. A. 
acknowledges the support by the Greek State Scholarship Foundation.

\end{document}